\documentstyle[12pt]{article} 
\newcommand{\be}{\begin{equation}}
\newcommand{\ee}{\end{equation}}
\newcommand{\beq}{\begin{eqnarray}}
\newcommand{\eeq}{\end{eqnarray}}
\newcommand{\bear}{\begin{array}}
\newcommand{\ear}{\end{array}}

\begin{document}
\title{ {\scriptsize{General Relativity and Gravitation, Vol. 31, No. 07 (1999),
1015-1030}}\\ \bigskip\bigskip
Decoherence in the Starobinsky Model} 
\author{S.Biswas $^{*a),b)}$, A.Shaw $^{**a)}$ and B.Modak$^{a)}$ \\
a) Department of Physics, University of Kalyani, West Bengal,\\ 
India, Pin.- 741235 \\
b) IUCAA, Post bag 4, Ganeshkhind, Pune 411 007, India\\
 $*$ e-mail : sbiswas@klyuniv.ernet.in\\
 $**$ e-mail : amita@klyuinv.ernet.in} 
\date{}
\maketitle
\begin{abstract}
Starobinsky described an inflationary scenario in which quantum corrections to
vacuum Einstein equations drive the inflation. The quantum cosmology of the
model is studied by solving the Wheeler-DeWitt equation. A connection between
uncertainty requirement, randomness in initial states and curvature
fluctuation is studied with a Schr\"{o}dinger-type equation through a time parameter
prescription. The result obtained is applied to understand the decoherence
mechanism in quantum gravity in the Starobinsky description.
\end{abstract}
\medskip
KEY WORDS : Fourth-order gravity
\section{\bf{Introduction}}
The Starobinsky scenario describes inflation through self-consistent solution
of vacuum Einstein equations,
\be
R_{\mu\nu}- {1\over 2} g_{\mu\nu}R ={ -8\pi G }< T_{\mu\nu}> ,
\ee
where $<T_{\mu\nu}>$ is the expectation value of the energy momentum tensor. With
a metric in Robertson-Walker form ,
\be
{ds}^2 = {dt}^2 -a^2 (t) {d{\sigma_k}}^2,
\ee
where ${d{\sigma_k}}^2$ is the metric on a unit three sphere for $k=1$, and 
the quantum
corrections take a particularly simple form for free, massless, conformally
invariant scalar fields:
\be
< T_{\mu\nu}> ={k_1}^{\;\;(1)}H_{\mu\nu} + {k_3}^{\;\;(3)}H_{\mu\nu},
\ee
where $k_1$ and $k_3$ are constants and
\be
{(1)_{_{H}}}_{_{\mu\nu}}=2R_{;\mu\, ;\nu}-2g_{\mu\nu}{R_{;\sigma}}^{;\sigma}
+2RR_{\mu\nu}-{1\over2}g_{\mu\nu}R^2
\ee
\be
{(3)_{_{H}}}_{_{\mu\nu}}={R_\mu}^\sigma R_{\nu\sigma}-{2\over 3}RR_{\mu\nu}
-{1\over 2}g_{\mu\nu}R^{\sigma\tau}R_{\sigma\tau}+{1\over 4}g_{\mu\nu}R^2
\ee
The characteristic features of the model are as follows:
\par
\noindent
(i) The model has a trace anomaly. Identifying
\be
8\pi G<T_{\mu\nu}>={H_o}^{-2\;\; (3)}H_{\mu\nu}+{1\over 6}{M_o}^{-2 \;\;(1)}H_{\mu\nu}
\ee
we get
\be
8\pi G<{T_\nu}^\nu>={H_o}^{-2}({1\over 3}R^2 -R_{\nu\sigma}R^{\nu\sigma})-M_{o}^{-2}{R_{;\nu}}^{;\nu}
\ee
though $<{T_\nu}^\nu>$ vanishes for the classical conformally-invariant fields.
\par
\noindent
(ii)The model shows de-Sitter solutions
\beq
a(t) & = & {H_o}^{-1} \cosh (H_ot),\;\;k=+1 \\
a(t) &= &{a_o} \exp (H_ot),\;\;k=0 \\
a(t)&=&{H_o}^{-1} \sinh (H_ot)\;\;k=-1
\eeq
These solutions describe an inflationary phase driven entirely by quantum corrections.
\par
\noindent
(iii)Studying instability it has been found [1] that the solutions (8) - (10)
are unstable under small perturbations and lead to a matter dominated universe at
late times with $ a(t)\propto t^{2\over 3}.$
\par
\noindent
(iv) The curvature fluctuations are found to be
\be
{({{\delta R}\over R_o})}^2\sim{ G{H_o}^2 \over \pi}.
\ee
The present work investigates the quantum cosmology of the
Starobinsky model. Vilenkin [2] attempted this problem in order to study 
tunneling from ``nothing''
to the Starobinsky inflationary phase. The quantum analysis is then used to calculate the
curvature fluctuations and assume a tunneling ansatz for the classical evolution of the model.
\par
In view of recent trends  
in solving the Wheeler-DeWitt equation [3-6], we take
up study of the Starobinsky description again to understand the curvature fluctuations, quantum 
force driving the Starobinsky inflation, initial conditions for the classical evolution from
quantum phase. We also study the quantum to classical transition through the mechanism of decoherence.
\par
It is worthwhile pointing out that, at the semiclassical level, the Starobinsky
inflationary scenario has been criticized [7] 
on the ground that the inflationary
solutions are not perturbatively expandable in the parameter of ``quantum 
corrections'' terms in the equation of motion or Lagrangian. There is also some 
question of treating the high order terms in the Lagrangian on an equal 
footing with the Einstein terms. Even then, studying the quantum cosmology of 
the Starobinsky description from the standpoint of quantum to classical transition 
would help us rethink the criticism of the model. Leaving 
aside this fact, it is instructive to look at the model as a toy example, to 
understand the current boundary condition proposals, decoherence mechanism and 
also the origin of quantum force in the early universe. Another motivation for studying
the Starobinsky description is due to its inbuilt mechanism leading to 
spontaneous nucleation of the universe. Among the various types of inflationary 
models, the choice of the inflaton potential has no microscopic origin, i.e. the 
choice of potential is not dictated from particle physics phenomenology (except
Coleman-Weinberg potential). For this reason, the warm inflationary scenario 
[8] has been a crazy inclusion among the various types of inflationary models. In
view of this fact, studying of the quantum cosmology of the Starobinsky model would 
be interesting, at least from the viewpoint of quantum to classical transition 
with a view to understand the initial conditions of the pre-inflationary era.
\par
In section II we obtain an approximate form of action for Starobinsky's  description 
thereby obtaining the Wheeler-DeWitt equation, and discuss some results of Vilenkin 
[2]
for comparison with our work. In section III we discuss the time prescription in
quantum gravity to reduce the Wheeler-DeWitt equation to a Schroedinger form containing time.
The initial conditions are then treated in the light of our discussion. In section IV,
we mainly concentrate on the decoherence mechanism to understand the quantum to classical transition
and this constitutes the main text of the present paper. In this paper we work 
with wormhole-dominance proposal [9] as a boundary condition to the 
Wheeler-DeWitt equation and find that Gaussian ansatz is the corresponding 
initial condition for the Schr\"{o}dinger-Wheeler-DeWitt equation. This 
correspondence is a breakthrough in the present trend of investigation along 
these lines [5,6]. Lastly, we end up with a discussion in section V.  
\section{\bf{Starobinsky Description}}
The evolution equation for the scale factor is obtained from (1) ,(4), and (5) for $ k = 0 $
and is given by
\be
{H^2(H^2 - {H_o}^2)} = {{{H_o}^2}\over {{M_o}^2}}( 2H\ddot{H} +6H^2 \dot{H} 
-\dot{H}^2)
\ee
In (12)$ H={\dot{a}\over a}$ and the dot denotes the derivative with respect 
to cosmological time, and
\be
H_o=({8\pi k_3 G})^{-{1\over 2}},\; M_o=({48\pi k_1 G})^{-{1\over 2}}.
\ee
Equation (12) gives de Sitter solution $ H=H_0 $. Equation (12) for the condition
$ \dot{H}< H^2 $ , $ \ddot{H} << H \dot{H}$ reduces to 
\be
H^2 - {H_o}^2 = 6 ({{H_o}^2\over {M_o}^2})\dot{H}
\ee
with the solution
\be
H=H_o\tanh({\gamma-{{{M_o}^2 t}\over 6H_o}})
\ee
where 
$\gamma = {1\over2}\ln({(H(o)-H_o)\over {H_o}})$. Long inflation requires

$M_{o}^{2}<<6H_{o}^{2}$. For $H<<H_o$ the term proportional to $H^4$ in
(12)
can be neglected.
\be
2H\ddot{H}+6H^2\dot{H}-{\dot{H}}^2+M_{o}^2H^2=0,
\ee
with an approximate solution
\be
H\simeq{4\over 3t}\cos^2 ({M_ot\over2})\left[1-{{\sin{M_ot}\over{M_ot}}} \right] \,.
\ee
Changing the origin of the time coordinates $M_ot\rightarrow(M_ot+\alpha)$, we get
\be
a(t)=const.\,{t^{2\over3}}\left[1+({2\over{3M_ot}})\sin{M_ot}\right]\,.
\ee
The expansion rate averaged over the oscillations period is
\be
\overline{H}={2\over 3t}
\ee
so that $\overline{a}\propto t^{2\over 3}$ corresponds to the matter 
dominated universe.
\par
We have carried out [10] numerical integration of (1) with a contribution from particle
production. We found that for a late time behaviour showing a radiation dominated
(and also for a matter dominated) evolution, the initial state emerges mostly
with a deSitter phase. Solutions (out of various solutions) with $a(t)=a_{min}$
at $t=0$ greatly effects both the initial and late time behaviour. This numerical
result suggests that the classical universe sits for infinite time at the turning
point $a=a_{min}$, but quantum mechanically it enters into the disallowed region
$a<a_{min}=H_{o}^{-1}$ and tunnels out again in the classical region with a 
curvature fluctuation $\sim {1\over{a_{min}}^2}\sim H_{o}^{2}$. To understand the 
quantum entanglement in the classical region, we need a quantum analysis in
minisuperspace description. We need an action for the quantum description to get
the Wheeler-DeWitt equation. For Starobinsky's description, (1) or (12), 
no closed
form of action is available. However for $M_o <<H_{o}^{2}$, a closed form of action
is obtained in the form
\be
S(R)= {24\pi \over G}({{1\over R}+{1\over {6M_{o}^{2}}}+{1\over R_{o}}
\ln{R\over R_o}})
\ee
with
\be
L(R)={1\over {16\pi G}}(R+{R^2\over {6M_{o}^{2}}}+{R^2\over R_o}\ln{R\over R_o})
\ee
where $R_o=12H_{o}^{2}$. The curvature fluctuation is
\beq
({\delta R \over R_o})^2 & \sim &{2\over {R_{o}^{2}\vert S''(R_o)\vert }}\nonumber \\
&\sim & {GH_{o}^{2}\over \pi}
\eeq
as expected.
\par
To obtain the Wheeler-DeWitt equation through canonical quantization, one needs to 
determine the Hamiltonian $H$ to obtain the equation
\be
H\psi=0
\ee
In view of higher order terms like $R^2$ and $R^2 \ln{R\over R_o}$ one gets 
\be
S=\int{\cal L}(a,\dot{a},\ddot{a})dt
\ee
in which the second derivatives $\ddot{a}$ cannot be removed from (24). 
To eliminate $\ddot{a}$ one 
writes
\be
S=\int{\cal L}(a,\dot{a},R,\dot{R})dt
\ee
and takes
\be S=2\pi^2 \int\{ L(R)a^3-\beta\left[ R-6a^{-2}(1+\dot{a}+a\ddot{a})\right]\}dt \,.
\ee
Varying $S$ with respect to $R$, we find
\be
\beta=a^3L'(R)\,.
\ee
Putting (27) in (26) one can remove $\ddot{a}$ through integration by parts. 
Introducing new variables $q$ and $x$ instead of $a$ and $R$,
\be
q=H_oa({L'\over {L'}_o})^{1\over 2}
\ee
\be
x={1\over2}\ln({L'\over {L'}_o})
\ee
and using the condition $M_{o}^{2}<<H_{o}^{2}$ one obtains the Wheeler-DeWitt
equation with the replacement $ {P_q}\rightarrow {-i}{\partial\over \partial q},
{P_x}\rightarrow {-i}{\partial\over \partial x }$ as
\be
\left[ {\partial^2 \over {\partial q^2}} - {1 \over q^2}{\partial^2
\over {\partial x^2}} - V(q,x) \right] \psi (q,x) = 0
\ee
In (30)
\beq
 x & = & {1 \over 2} \ln ({R \over R_o})  \\
V(q,x) & = & \lambda^{-2} q^2 (1 - q^2 + \mu^2 (x) q^2)  \\
\mu^2(x) & = & {{M_o^2} \over {2{H_o}^2}} (2x + e^{-2x} -1)  
\eeq
where
\be
\lambda = {{G{M_o}^2} \over {6 \pi}}\,.    
\ee
We make the transformation
$Q = {q \over {\sqrt{\lambda}}}$, 
in (30) and get
\be
\left[ {{\partial^2} \over {\partial Q^2}} - {1 \over Q^2}{{\partial^2}
\over {\partial x^2}} - Q^2 (1 - Q^2 U(x)) \right] \psi = 0
\ee
where
\be
U(x) = \lambda (1 - \mu^2 (x))\, .
\ee
Vilenkin obtained a solution of (35) in the region $q>q_x$, where 
$q=q_x\simeq 1+{1\over 2}\mu^2(x) $ is the turning point,
\be
\psi =\pi^{1\over 2}{\vert V_o(q)\vert}^{-{1\over 4}} 
\exp { \{ -{1 \over {3 \lambda}} (1 + \mu^2 (x)) \left[ 1 - 
i (q^2 - q^2 \mu^2 -1)^{3/2} \right] +{i\pi\over 4}\}}
\ee
where 
\be
V_o(q)=\lambda^{-2}q^2(1-q^2)\,.
\ee
We take $q>>1$ in (37) and find
\beq
\psi & \simeq & \pi^{1\over 2}\lambda^{1\over 2}q^{-1} 
\exp { \{ -{1 \over {3 \lambda}} (1 + \mu^2 (x)) \left[ 1 - 
i q^3 (1- \mu^2(x))^{3/2} \right] +{i\pi\over 4}\}} \nonumber \\
& \simeq & (\pi\lambda)^{1\over 2}q^{-1} 
\exp { \{ -{1 \over {3 \lambda}} (1 + \mu^2 (x)) \left[ 1 - 
i q^3 (1- \mu^2(x))^{3/2} \right] \}}\,. 
\eeq
For $x<<1$,
\be
{\vert \psi \vert}^2 \propto \exp { \left[ -{{4 \pi x^2} \over
{G{H_o}^2}}
\right]}\,.
\ee
If one temporarily sets aside the normalization restriction on $\psi$,
\be
< {({{\delta R} \over {R_o}})}^2> = 4 <x^2> \sim {{G{H_o}^2} \over {2
\pi}},
\ee
in qualitative agreement with the classical result.
The fluctuation in expansion rate is
\be
\delta_o\equiv{\delta H\over {H_o}}={1\over 2}{\delta R\over R_o}
\sim ({{GH_{o}^{2}}\over {8\pi}})^{1\over 2}\,.
\ee
This $\delta_o$ determines the duration [see (15)] of the inflationary phase in the
model
\be
t_*\sim ({3H_o\over {M_{o}^{2}}})\ln{2\over \delta_o}\,.
\ee
Taking $H_o\sim 0.7 m_p$ and using (42) and (43) we get $\delta_o\sim 0.14,\,
t_*\sim 8{H_o\over {M_{o}^{2}}}$. Using $8\pi k_1=1.8$ (restriction from $SU(5)$
model) and the value of $M_o$ from (13),
$t_*\sim 4\times {10}^{10}H_{o}^{-1}$. This value of $t_*$ is more than 
sufficient to solve the horizon and flatness problem. The results quoted so far 
are standard and will be needed to understand much of what follows. 
\section{\bf{Reduction to Schr\"{o}dinger Form}}
Recent trends in quantum cosmology [3-6] suggest that the solution of (35)
should be written as 
\be
\psi(Q,x)=\exp{\left[iS(Q)\right]}\Phi(Q,x)
\ee
in which WKB action $S$ depends only on $Q$. However the philosophy behind this 
reduction though differs, most of the workers obtained Schroedinger equation 
through a prescription on time variable. Apart from the interpretation, almost
everybody remains silent about the normalization, and consequently also about the
probability concept. These two concepts are of utmost importance for the 
interpretational framework of the wavefunctional $\psi(Q,x)$. We have shown [9] 
that if one takes contribution from wormholes in WKB ansatz, the 
wavefunctional becomes normalizable. When boundary conditions corresponding to 
`tunneling' and `no boundary' proposals are introduced, one recovers the respective
wavefunction. We have shown elsewhere that if one introduces the probabilistic 
concept through a continuity-like equation which also gives a time parameter
prescription, the wavefunctional $\psi(Q,x)$ separates into the form (44)
provided the classical Hamilton-Jacobi equation remains satisfied. We have thus
a dynamical content to the form (44). We report briefly here the results 
only [11,12]. 
We put (44) in (35) and use        
\be
{\partial \over {\partial t}} = - {1 \over Q}{{\partial S} \over {\partial Q}}
{{\partial} \over {\partial Q}}
\ee
and we find
\be
i{{\partial \Phi} \over {\partial t}} = \left[ {-{1 \over {2Q^3}} 
{{\partial^2} \over {\partial x^2}}} + {Q^3 \over 2}U(x) + {1 \over {2Q}} 
\right] \Phi\,.
\ee
Remembering that ${\partial^2 \over {\partial Q^2}}$ in (35) remains multiplied 
by a factor $\hbar^2\, and \,S$ in (44) $\sim {S\over \hbar}$, we neglected terms
${{\partial^2 \Phi}\over {\partial Q^2}}$ which are of the order of $\hbar^2$. It
may be noted that our prescription does not violate unitarity which is obtained
in the other prescription. The assumption (45) is consistent with (35) as well as
with the Einstein classical equation. Equation (46) is the time dependent Schroedinger equation 
in quantum gravity.
\par
To understand the salient features of the Starobinsky description, let us proceed with
the solution of (46) with Gaussian ansatz
\be
\Phi (Q,x) = N (t) e^{- {{\Omega (t)} \over 2} x^2}.
\ee
Substituting (47) in (46) we get coupled equations
\be
i{d \over {dt}} \ln N = {\Omega \over {Q^3}} + {Q^3 \lambda}  
+ {1 \over Q}
\ee
\be
i {\dot{\Omega}} = {\Omega^2\over {Q^3}} 
+ {{M_{o}^{2}\lambda} \over H_{o}^{2}}Q^3.
\ee
With the anstaz
\be
\Omega = - i Q^3 {{\dot{y}} \over y}
\ee
one finds from (49)
\be
{\ddot{y}} + 3 {{\dot{Q}} \over Q}{{\dot{y}}} 
- {{\lambda M_{o}^{2}}\over {H_{o}^{2}}} y = 0.
\ee
Introducing conformal time coordinate $\eta $ according to $ dt = Q d\eta $, one finds
\be
y'' + 2 {{Q'} \over Q} y' - {{\lambda{ M_o}^2 Q^2}\over { H_o}^2 }y = 0
\ee
where prime denotes a derivative with respect to $\eta $. With the view that the classical
model has a de Sitter solution , we take 
$ Q (\eta ) =-{1\over \sqrt{\lambda}\;\;\eta }$,
where $\eta$ runs from $-\infty\, to\; 0 $.
Equation (52) now reads
\be
y'' - {2 \over \eta} y' - {{M_o}^2 \over{ H_o}^2\eta^2} y = 0
\ee
which is solved by
\be
y = \eta^{{3/2} \pm {\sqrt{{9/4} + m^2}}}.
\ee 
where we take $m^2={{M_0}^2\over{ H_0}^2}$. In the limit $m^2 << {9\over 4}$
(which is usually assumed to be satisfied in the inflationary model and in
our description) such that
\be
\sqrt{{9/4} + m^2} \sim ({3/2} +{{m^2}/3}) 
\ee
In the conformal time, the expression (50) reads
\be
\Omega = -i Q^2{y'\over y}
\ee
so that using (54) and (55) one finds
\be
\Omega = -i m^2 \sqrt{\lambda} {Q^3 \over 3}.
\ee
As $\Omega$ is imaginary the state (47) is not normalizable. To study decoherence
one needs a real part in (57). We take up decoherence in the next section.
\section{\bf{Decoherence mechanism}}
Decoherence is a mechanism through which we understand how the classical world arises
from a quantum wavefunction of the universe. In decoherence, quantum  interference
effects are suppressed by the averaging out of microscopic variations not 
distinguished by the associated observables. In standard quantum theory this is referred
to as the `collapse of the wavefunction'. It is a formidable task to forbid the
occurrence of linear superposition of states localized in far away spatial regions
and induce an evolution agreeing with classical mechanics. In the context of quantum
gravity the situation further complicates due to the absence of "time" 
because the Wheeler DeWitt
equation when compared to Schr\"{o}dinger equation
\be
i\hbar{{\partial\psi}\over \partial t}= H\psi,
\ee
gives a timeless character to the wavefunction of the universe. Though we 
recovered the form
(58) with a reduced Hamiltonian given in (46), it must be ensured that during the
quantum to classical transition , none of the successful quantitative predictions of the
inflationary scenario for the present day universe is changed. An important
aspect in this direction is to choose an initial condition for eqs. (58) or (35)
such that basic input ``inflation'' remains undisturbed in the description.
With respect to the Wheeler-DeWitt equation, the wavefunction $\psi$ is described with
some boundary condition proposals namely (i) Hartle-Hawking 
[13] (ii) Vilenkin [14]
and (iii) Wormhole-Dominance proposal [9]. The latter one is recently proposed by us.
\par
Now we will show that an initial adiabatic ground state with a Gaussian form
is a suitable choice provided the quantum cosmological initial conditions
correspond to the wormhole dominance proposal, or at least to the Hartle-Hawking
proposal. This would also justify the correct choice of the boundary conditions.
Using (57) in (48), we obtain for large $Q$
\be
N = N_o \exp { \left[ - {{i Q^3 \sqrt{\lambda}} \over 3} \right]}.
\ee
The constant $ N_0$ will be evaluated through the wormhole dominance
proposal.
The wavefunction now reads, corresponding to (46) or (58),
\be
\psi = N_o \exp { \left[- {{i Q^3 \sqrt{\lambda}} \over 3} 
+ i{{M_{o}^{2} \sqrt{\lambda}Q^3} \over {6H_{o}^{2}}} x^2\right] }\,.
\ee
We write the exponent in (60) as
\beq
S & = & - {{i Q^3 \sqrt{\lambda}} \over 3} 
{(1-{{1\over 2}{M_{o}^{2}\over {H_{o}^{2}}} }}x^2) \nonumber \\
& = & -{i \over 3} {\left[ {Q^2 \lambda}(1 -{ M_{o}^{2}\over {H_{o}^{2}}}x^2) \right]}^{3/2} 
{1\over {\lambda (1-{M_{o}^{2}\over {H_{o}^{2}}}x^2)}}
\eeq
since ${M_{o}^{2}\over H_{o}^{2}}x^2<<1$. Comparing with (33) we write (61)
\beq
S&=& -{i\over {3\lambda(1-\mu^2(x))}}
\left[ \lambda Q^2(1-\mu^2(x))\right]^{3/2}\nonumber \\
&=& -{i\over {3\lambda(1-\mu^2(x))}}
\left[ q^2(1-\mu^2(x))\right]^{3/2}\nonumber \\
&\simeq & -{i\over {3\lambda}}(1+\mu^2(x)) 
\left[ q^2(1-\mu^2(x)) -1 \right]^{3/2}\,.
\eeq
The last step will be justified from the Wheeler-DeWitt equation itself since 
$q>>1$. The wormhole dominance proposal considers incorporating the repeated
reflections from the turning points $q=o$ and 
$q=q_x={1\over {(1-\mu^2(x))^{1\over 2}}}$ to contribute to $N_o$ as
\be
N_o = {{\exp {S(q_x,0)}} \over 
{1 - \exp {\left[ 2 S(q_x,0) \right]}} }\,.
\ee
Here
\be
S(q_x, 0)\equiv \vert S\vert_{0}^{q_x}\,. 
\ee
Evaluating (63) we find
\be
N_o= {\exp{\{{1 \over {3\lambda}}(1+\mu^2(x))\}}
\over {(1- \exp{\{{2/{3\lambda}}}(1+\mu^2(x))\})}}
\ee
Hence
\be
\psi= {\exp{\left[{1 \over {3\lambda}}(1+\mu^2(x))
\{1-i(q^2(1-\mu^2(x))-1)^{3\over 2}\}\right]}
\over {(1- \exp{\{{2/{3\lambda}}}(1+\mu^2(x))\})}}\,.
\ee
We continue (66) in the region $q^2(1-\mu^2)<1$ to get
\be
\psi={\exp{\{{1 \over {3\lambda}}(1+\mu^2(x))
\left[1-{(1-q^2+q^2\mu^2)}^{3/2}\right]\}}
\over {(1- \exp{{2/{3\lambda}}}(1+\mu^2(x)))}}\,.
\ee
If we leave aside the denominator of (67), the wavefunction (67) corresponds to
the Hartle-Hawking wavefunction.
\par
Thus the initial condition for the Schr\"{o}dinger-Wheeler-DeWitt equation (46) 
turns out to be
that at an early time (near the onset of inflation), the modes are in their
adiabatic ground state due to wormhole dominance and this serves as the initial
conditions on the Wheeler-DeWitt wavefunction. Unlike the claim by some authors,
it also establishes that Hartle-Hawking initial conditions also provides inflation
and serve as a seed for decoherence to suppress the interference for quantum to 
classical transition. One important feature of (67) is that at $q\rightarrow 0$,
(67) reduces to  
\be
\psi\sim {{\exp\{{q^2\over {2\lambda}}\}}
\over {(1- \exp{\{{2/{3\lambda}}}(1+\mu^2(x))\})}}\,,
\ee
characteristic of the Hartle-Hawking wavefunction. As is evident from (34), 
$\lambda$ 
is a small quantity, and hence, considering the denominator of (68), we
find
\[ \vert\psi\vert^2\sim e^{-{1\over {3\lambda}}(1+\mu^2(x))}\,\]
This gives the curvature fluctuation as in (41).
\par
The steps leading to the form (68) from (60) indicates that at the very early stage 
(within the turning points) the quantum superposition principle has been effective
such that the time has lost its meaning. The interference between $e^{iS}$ and
$e^{-iS}$ is maintained through the wormholes that act as a driving quantum 
force. We have shown elsewhere [15] that without any reference to the Wheeler-DeWitt
equation one can also recover eq. (46) exactly  
using only the classical Einstein-Hamilton-Jacobi equation with a directional
time derivative
\be
{\partial\over {\partial t}}=-{1\over Q}{{\partial S_o}\over {\partial Q}}
{\partial\over {\partial Q}},
\ee
i.e., the scale factor (gravitational field) itself acts as time. Just at the classical
turning point, time begins to flow and the effect of the quantum turning point 
(at which the wormhole contributes) only survives through $N(t)$ in the Gaussian
ansatz i.e., the states $e^{iS}$ and $e^{-iS}$ begin to decohere. It has been 
guessed by some authors [16] that some sort of boundary 
conditions at small scales
may lead to quantum effects in the vicinity of the turning point. The `wormhole
dominance' proposal exactly establishes this aspect, keeping the necessary coinage
for inflation. If we are to bridge the solution (47) with (67), the reduction
\be
S=S_o(Q)+S_1(Q,x)
\ee
with
\[\psi=\exp{\left[ iS_1(Q,x)\right]}\]   
is an unavoidable fact. Kiefer [5] named this reduction as the `relevant' and the
`environmental' degrees of freedom. The solution (67) refers to the ground state 
wavefunction. The higher degrees of freedom (`higher multipoles') when taken
into account, eq. (53) modifies to  
\be
y''-{2\over \eta}y'+({n^2-{{M_{o}^{2}}\over {H_{o}^{2}\eta^2}}})\;y=0\,.
\ee
Evaluating $\Omega$ along the lines of [5] we find
\be
\Omega\simeq {{n^2Q^2}\over {n^2+Q^2H^2}}(n+iQH_o)-i{{M_{o}^{2}Q^3}\over 
{3H_o\lambda}}\,,
\ee
and the decoherence factor responsible to suppress interference terms is written 
as
\beq
\exp{(-D_{\pm})}&=&\exp{(-{1\over 4}Tr\,{{\Omega_{I}^{2}}\over 
{\Omega_{R}^{2}}})}\nonumber \\
&=&\exp{\sum_{n}n^2({\Omega_I\over \Omega_R})^2}\,.
\eeq
\par
We choose the adiabatic vacuum state with positive frequency as the solution 
of (71) or (53) and this acts as initial vacuum state. This initial state evolves
at $\eta\rightarrow +\infty$ according to a WKB form,
\be
y={\alpha\over {\sqrt{2\omega}}}e^{i\int^{\eta}\omega(\eta')d\eta'}
+{\beta\over {\sqrt{2\omega}}}e^{-i\int^{\eta}\omega(\eta')d\eta'}\,,
\ee
where $\alpha$ and $\beta$ are the usual Bogolubov co-efficients. For the
WKB
state (74) one finds 
\beq
\Omega&\simeq&\omega+{{i\dot{\omega}}\over {2\omega}}\\
&=&\sqrt{n^2-m^2Q^2}-i{{m^2QQ'}\over {2(n^2-m^2Q^2)}}\,,
\eeq
with $m^2={M_{o}^{2}\over H_{o}^{2}}\lambda$. Hence
\be
D_{\pm}={{m^4Q^2{Q'}^2}\over 8}\sum_{n}{n^2\over {(n^2-m^2Q^2)^3}}\,.
\ee
Using
\be
{1\over \zeta}-\pi \cot{\pi \zeta}=\sum_{1}^{\infty}{{2\zeta}\over
{n^2-\zeta^2}}
\ee
and after some straightforward manipulation we find 
\be
D_{\pm}={{\pi m^3QQ^{'2}}\over {128}}\left[{\pi\over {mQ}}
(1+\cot^2 \pi mQ)+{{\cot\pi mQ}\over {(mQ)^2}}
-2\pi^2\cot\pi mQ(1+\cot^2\pi mQ) \right] \,.
\ee
In the limit $mQ>>1$, one finds
\be
\exp (-D_\pm)\simeq\exp (-{{\pi m Q^{'2}}\over {128Q}})
=\exp (-{{\pi m H_{o}^{2}Q^3}\over {128}})\,.
\ee
This result is also obtained by Kiefer [6]. 
For $m\simeq 100\,GeV,\;H_o\simeq
55\,{km}/{sec\,Mpc} $, and $Q\simeq H_{o}^{-1}$, this is 
\[ \exp (-{10}^{+43})\simeq 0 \,.\]
Hence decoherence is efficient for large $Q$ even in the Starobinsky
description.
This argument leads us to reconsider the Starobinsky 
description despite the criticisms labelled against it.
\section{\bf{Discussion}}
The Starobinsky model provides a viable description both in the classically allowed
and in the forbidden region. It is a model where one loop quantum corrections 
spontaneously allow the universe to make the transition to a deSitter phase and then
to a classical Friedmann stage with a small perturbation. It is a model of
spontaneous nucleation rather than inflation-driven inflation. In our previous work
[9] we achieved the advantage of normalized wavefunction with probabilistic
interpretation but there still remains no certainty about the 
adoption of a particular boundary
condition proposal. Within the perspective of the 
Schroedinger-Wheeler-DeWitt equation describing the time evolution of the
universe, we found that Hawking type boundary conditions definitely establish
inflation (since we take $Q=-{1\over {\sqrt{\lambda}\;\,\eta}}$), as the characteristic
feature of Starobinsky model, unlike the claim by some workers in favour of the
tunneling proposal. The wormhole dominance proposal serves as an initial condition 
for decoherence to effect quantum to classical transition, because
the Gaussian
ansatz for SWD equation (valid in the classically allowed region) leads, after
continuation, to the wavefunction according to the wormhole dominance proposal.
We observe that the Starobinsky model (basically reminding us of a $R^2$-cosmology)
is equivalent to an Einstein gravity plus a scalar field and this
inflaton-like scalar field arises from one loop quantum corrections. 
It may be mentioned that quantum cosmology with Hartle-Hawking boundary 
conditions for a model
Einstein gravity plus a dialaton was studied by Okada and Yoshimura [17].
They briefly comment that the nucleation of the classical universe would be 
exponentially suppressed if one takes the tunneling proposal. The quantum 
nature of the early universe thus guarantees the universal validity of 
superposition principle. This in turn incorporates multiple reflections between
the turning points. The wormhole dominance proposal takes this fact justifiably
defining a normalization constant on the basis of superposition principle
favouring probabilistic interpretation rather than the concept of `conditional
probability'. In our previous work [9] we have shown that the normalization 
constant prescription in the wormhole dominance proposal is equivalent to
the contribution of wormholes leading to an interpretational framework for
wormhole picture (lack of which created a confusion about its introduction),
apart from Klebanov and Coleman's arguments, as a driving quantum force in the
early universe. The ensemble of quantum universe models would thus be a realistic
situation in quantum gravity.    
\smallskip
\section{\bf{Acknowledgment}}
A. Shaw acknowledges the financial support from ICSC World Laboratory,
LAUSSANE, Switzerland during the course of the work.
\smallskip
\begin{center}
{\bf{References}}
\end{center}
{\obeylines\tt\obeyspaces{ 
1. Vilenkin A.,(1985), Phys. Rev. D{\bf{32}}, 2511. 
2. Vilenkin A.,(1983), Phys. Rev. D{\bf{27}}, 2848. 
3. Wada S.,(1986), Nucl. Phys. B{\bf{276}}, 729.
4. Kuchar K.V.,(1992), In Proc. of the fourth Canadian 
\quad Conference on General Relativity and Relativistic  
\quad Astrophysics, edited by G. Kunstatter, D. Vincent and
\quad J. Williams (World Scientific, Singapore), P 211-314.
5. Kiefer C.,(1992), Phys. Rev. D45, 2044.
6. Kiefer C.,(1992), Phys. Rev. D46, 1658.
7. Simon J.Z.,(1992), Phys. Rev. D45, 1553
\quad Parker L. and Simon J.Z.,(1993), Phys. Rev. D47, 1339.
8. Berera A.,(1997), Phys. Rev. D55, 3346 and ref. there in.
9. Biswas S.,Modak B. and Biswas D.,(1996),Phys. Rev. D55, 4673.
10. Shaw A., Biswas D., Modak B. and Biswas S.,(1999) 
\quad Pramana - J. Phys. 52, 1. 
11. Biswas S., Shaw A. and Modak B.,(1998), "Time In Quantum 
\quad Gravity", gr-qc/9906010.    
12. Biswas S., Shaw A. and Biswas D.,(1998), "Schroedinger 
\quad Wheeler-DeWitt Equation In Multidimensional Cosmology", 
\quad gr-qc/9906009.   
13. Hawking S.W.,(1984), Nucl. Phys. B239, 257
\quad Hawking S.W. and Page D.N.,(1986), Nucl. Phys. B264, 184.
14. Vilenkin A.,(1987), Phys. Rev. D37, 888.
15. Biswas S., Shaw A., Modak B. and Biswas D.,(1998), "Quantum  
\quad Gravity Equation In Schroedinger Form", gr-qc/9906011. 
16. Conardi H.D. and Zeh H.D.,(1991), Phys. Lett. A154, 321.
17. Okada Y. and Yoshimura M.,(1986), Phys. Rev. D33, 2164.
}} 
\end{document}